# Broadband THz emission of long pulses from photomixing process with optical chirped pulses.

GABRIEL TATON,[1,*] FREDERIC FAUQUET,[1] ILYES BETKA,[1] JEAN-PAUL GUILLET,[1] FREDERIC DARRACK,[1] PATRICK MOUNAIX,[1] DAMIEN BIGOURD[1]

[1]*Laboratoire IMS, UMR CNRS 5218, Université de Bordeaux, 33400 Talence, France*

*Corresponding author: gabriel.taton@u-bordeaux.fr

**Terahertz (THz) generation via photomixing on photoconductive antenna using twin delayed chirped pulses provides a long THz pulse with a narrow bandwidth. To generate a long pulse with broad bandwidth, we propose a new method that combines two long optical pulses with opposite chirps. The pulses exhibit temporal distributions of their instantaneous frequencies with opposite slopes. As a result, interaction between the beat frequency evolving over time and a photoconductor produces a broad THz spectrum with temporal variations. In our experimental setup, we generate a 12 ps-long pulse with a 1 THz bandwidth spectrum, featuring a frequency ramp of 90 GHz/ps, resembling a chirped THz pulse. This approach signifies a major advancement toward integrating photomixer technology, particularly in THz ranging applications.**

The terahertz (THz) domain, which lies between microwaves and infrared (0.1 to 10 THz), offers numerous ground-breaking applications, particularly in telecommunications [1], spectroscopy [2,3], nondestructive testing and imaging [4, 5]. The field of THz spectroscopy has seen significant progress in enhancing pulse energy, tunability, and bandwidth. THz pulses can be generated through the down-conversion of ultra-fast laser pulses via nonlinear interactions in various materials. Besides, THz systems have been developed using ultrafast photoconductors, where a semiconductor is integrated with a metallic THz antenna. When the photon energy exceeds the semiconductor's band gap, mobile electron-hole pairs are generated and accelerated by biased electrodes, resulting in a transient current oscillating at THz frequency, which produces the radiation. Femtosecond lasers are commonly employed as the optical pump source in combination with photoconductor, to generate sub-ps broadband THz pulses. For continuous wave (CW) operation, the photomixing process is akin but it usually involves two CW lasers that are superimposed and focused on the photoconductor, known as a photomixer, with a beatnote that falls within the THz range [6-10]. This method offers high frequency resolution for frequency domain spectroscopy determined by the linewidth of the two CW lasers. The frequency scan is usually achieved by stepwise tuning one of the two lasers [11-13]. One CW laser can also be frequency-swept relative to a static laser for fast spectral scanning, using a scheme based on linearly frequency-modulated CW (FMCW) terahertz radiation [14,15].

Alternatively, by using ultrafast lasers, a quasi-CW regime can be achieved by stretching (or chirping) the pulse to the 100 ps or nanosecond scale. This temporally spreads the instantaneous frequencies, making their manipulation crucial for THz photomixing with high-power ultrafast lasers [16-18]. The use of chirped and delayed pulses provides a method for producing narrow band beat note source at a frequency ⍵THz from a broadband laser pulse, achieved by combining two pulses with a time delay ΔT (Figure 1.a). Tunability is thus achieved by scanning this relative time delay ΔT between the two pulses. In this configuration, the spectral resolution is obviously reduced to tens of GHz compared to the standard photomixing approach due to the characteristics of the laser. On the other hand, chirping ultra-short pulses reduces peak power, enabling antennas multiplexing [19,20] in an integrated device, thereby facilitating high average power at THz frequency [21-23]. However, in this configuration, reducing the peak power with chirped pulses results systematically in a narrow spectrum. In this manuscript, we propose an innovative method to deliver long pulses at THz frequencies with a broad THz bandwidth by interfering two optical pulses with opposite chirps (Figure 1.b). These pulses exhibit temporal distribution of their instantaneous frequencies with opposite slopes. As a result, the frequency beating evolves over time, and interaction with a photoconductor produces a broad THz spectrum with both spectral and temporal characteristics. In this scenario, the beat note frequency varies temporally within the pulse envelope, causing the emitted THz spectrum to follow a THz frequency ramp, similar to the FMCW technique but on a timescale of tens of picoseconds. In this work, we present an experimental configuration of the concept and compare the photomixing process for a narrow and broad band emission.

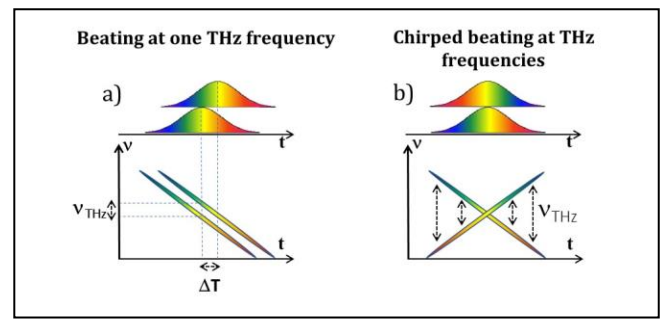

*Figure 1: Didactic scheme for narrow (a) or broad band (b) THz emission from chirped pulses.*

Figure 2 illustrates the experimental setup, which included three optional configurations. The laser emitted ultrashort pulses at a repetition rate of 76 MHz, with a pulse duration of 70 fs and a spectrum centered at 1028 nm. Both the THz emitter and detector were based on GaBiAs photoconductive devices optimized for a 1 µm optical excitation wavelength [24]. To determine the maximum THz bandwidth achievable with the set-up, the first configuration directly used the ultra-short pulse to excite the emitter. In the other two configurations, twin pulses were produced with either identical or opposite chirps. In the former case (box-a in Figure 2), the optical pulse passed through one side of a volume Bragg grating (VBG), introducing a chirp with a linear rate of ~6.2 ps/nm [25] centered at 1030 nm, stretching the pulse from a duration of 70 fs to 93 ps (at Full Width at Half Maximum). The VBG provided a linear chirp rate to prevent any nonlinear frequency-time distribution of the THz radiation due to high order dispersion terms [26]. After reflection, the pulse was transmitted through a 10 nm band-pass filter. This spectral width, corresponding to 2.8 THz, set the maximum tunability achievable by the photomixing. Then, the chirped pulse was split within a Michelson interferometer, leading to two delayed pulses directed toward the THz emitter, with a total optical power of up to 25 mW. The radiated THz beam was collected with a silicon hemi-sphere lens and four 2-inch diameter off-axis parabolic mirrors. The signal was collimated and then focused into the THz detector. Standard photoconductive sampling was accomplished by adjusting a motorized delay stage to temporally overlap the optical probe pulse with the THz beam. The probe pulse was originated directly from the laser, ensuring an ultrafast response with a sub-ps resolution. In this set-up, photomixing was achieved with two delayed pulses that had the same chirp, resulting in a constant beat frequency determined by the interferometer's delay. Figure 3.a shows a THz temporal trace emitted when the beat frequency was set at 0.5 THz, with a bias voltage of 15 V and an optical power of 17 mW. The pulse duration of the electric field was 54 ps (FWHM). For comparison, the THz pulse excited with the ultra-short pulse had a duration of 1.7 ps. The corresponding spectrum, shown in Figure 3.b (blue line) has a bandwidth of ~18 GHz and is narrower than the spectrum of a THz pulse emitted from the antenna excitation with a single ultra-short pulse (red line). As the relative delay is scanned, the THz spectrum can be tuned up to ~1.3 THz as seen in Figure. 3.b (black dotted lines). The intensity of the main frequency follows the shape of the short-pulse spectrum, indicating that the limitation in frequency tunability is determined by the antenna's response and the dynamics of the detected field. Additionally, weaker peaks, separated by approximately 150 GHz, can be observed around the main frequency component. These are likely attributed to a Fabry-Pérot effect in the antenna, generating weaker pulses delayed by few picoseconds.

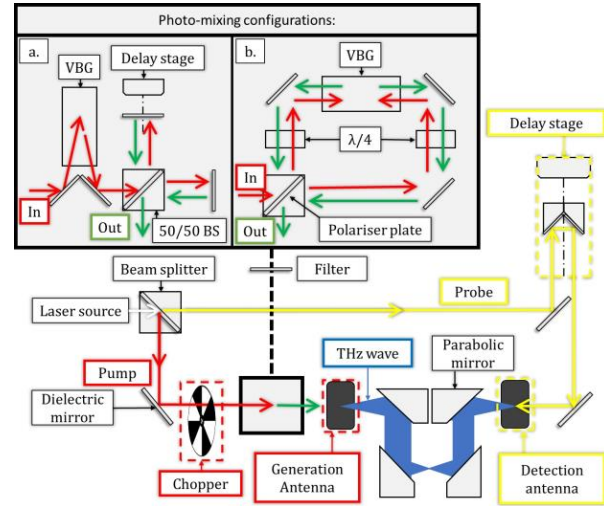

*Figure 2 Experimental set-up with 3 cases when the emitter is excited with an ultra-short pulse, 2 pulses with the same chirp (box-a) or 2 pulses with opposite chirps (box-b).*

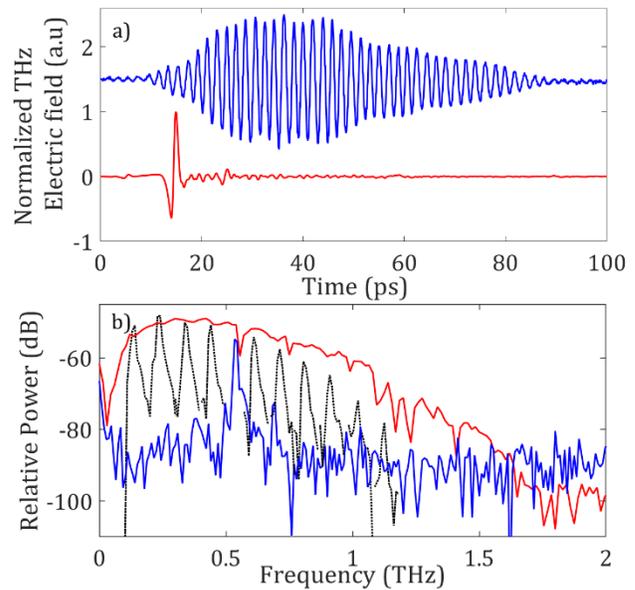

*Figure 3: a) THz electric field temporal shape for an excitation with (red) an ultra-short pulse or (blue) delayed chirped pulses with a beating frequency set at 0.5 THz. b) Corresponding spectra. The black dotted lines show the tunability when the delay between the two pulses is scanned.*

To increase the spectral bandwidth of the THz pulse while maintaining a long THz pulse, we used the second configuration (box-b in Figure 2) which involves two optical pulses with opposite chirps. The pulse was initially split in two arms, with each pulse reflected off different sides of the VBG. As a result, the two pulses were chirped to the same duration, but with opposite chirp slopes. Consequently, the beat frequency evolved linearly over time, with high frequencies at the edges of the overlap and low frequencies at its center (Figure 1.b). Through photomixing on an antenna, this leads to the generation of a long THz pulse with a symmetrical electric field around the pulse center. The optical power and bias voltage were set at 15 mW and 25 V, respectively. The oscillating pattern clearly indicates that the carrier frequency evolves over time (Figure 4.a). As the beat frequency varies within the pulse envelope, the emitted THz spectrum follows a frequency ramp, resembling a chirped THz pulse. Notably, the lowest frequency is emitted in the center of the pulse at ~15 ps, while the higher frequencies are generated at the leading and trailing edges. This results in two identical THz frequencies being generated at different times, leading to spectral interference, observed in the fast Fourier transform spectrum of the entire electric field (Figure 4.c). For comparison, half of the temporal trace was numerically selected and its spectrum aligns with the modulated spectral profile (Figure 4.c). The spectrum spans up to approximately 1 THz, which is lower than that achieved with the twin chirped pulses. Similarly, the temporal trace is shorter, around 12 ps, confirming that the high frequencies at the leading and trailing edges were not detected. This is likely due to a low dynamic range that limits their detection. The THz electric field was also analyzed using a 4 ps temporal gate, which was numerically slid across the entire temporal trace. At each step, a fast Fourier transform was performed, resulting in a spectrogram (Figure 4.b). This spectrogram confirms that the THz frequencies are linearly distributed over time with a slope of approximately 90 GHz/ps. The spectrogram is also compared to the one generated with twin delayed pulses (Figure 4.e), where as expected, a single frequency is emitted throughout the entire temporal trace.

Next, we examined the dependence of the DC photocurrent in the emitter and the emitted electric field on various experimental parameters. At relatively low values, the photocurrent increases linearly with respect to both the power and bias voltage (Figure 5), confirming ohmic transport, as expected in ideal conditions. However, as the power increases, the behavior changes due to effects such as carrier screening and charge transport [27]. The photocurrent remains approximately the same for both photomixing configurations, although at higher bias voltage and power, it is higher for the twin pulses. Similarly, the photocurrents generated with ultrashort pulses are within the same range even so the carrier has different dynamics in all configurations. The chirped pulse duration is much longer than the carrier lifetime. This means that the optical excitation persists over several carrier recombination cycles [28]. During these temporal intervals, the carriers oscillate at a constant frequency and their contributions can coherently add up, potentially enhancing the narrowband THz emission. In the case of large bandwidth emission, the behavior of photocarriers is significantly different. During each interval, the carriers oscillate at different instantaneous frequency, meaning that each spectral component is emitted over a shorter interval. Additionnaly, we should note that the oscillating carriers are responsible for the THz emission which is not systematically reflected by the measured DC photocurrent. Figure 5.b shows the maximum peak-to-peak electric field amplitude $E_{p-p}$ as the function of optical power P for optical chirped pulses. The electric field amplitude is significantly lower in the case of opposite chirps although the DC photocurrent is approximately the same. In addition, while a linear behavior is expected at low optical power, the electric field amplitude quickly saturates, even at low power levels. From the fit (dashed line in Figure 5.b) with the function $E_{p-p} \propto P/(P + P_{sat})$, the power $P_{sat}$, where the saturation occurs, can be extracted. For example, at 25 V, saturation is reach at $P_{sat}$= 2.7 mW for pulses with opposite chirps. In contrast, when the two pulses have an identical chirp, saturation occurs at a higher power with $P_{sat}$=8.5 mW. This strong saturation leads to a significant reduction in the THz electric field, by approximately a factor of 4, between the two cases.

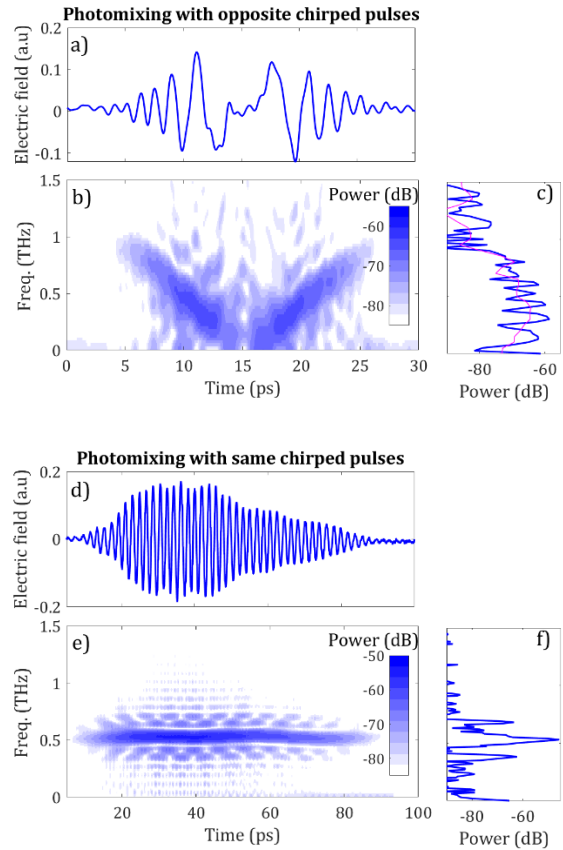

*Figure 4: Electric field emitted by long pulses with opposite chirps (a) or an identical chirp at 500 GHz (d) with their corresponding spectra (c,f) and spectrograms (b,e) in the logarithmic scale. The optical power and voltage are respectively for a power of 15 mW/25 V and 17 mw/15 V. The spectrum in magenta (c) is obtained from half of the THz waveform.*

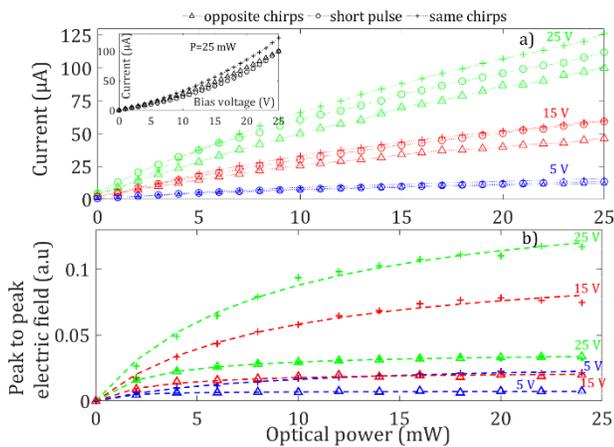

*Figure 5: Measurement of the photocurrent as the function of the optical power for various configurations. The inset corresponds to the photocurrent as the function of the bias voltage (b) Measurement of the peak-to-peak electric field as the function of the optical power for chirped pulses.*

THz generation via photomixing on a photoconductive antenna using long pulses with opposite chirps successfully produces a long THz pulse with a broad bandwidth and a linear frequency-time distribution. This innovative method marks a substantial step toward the integration of photomixer technology, as it allows both photoconductive and optical components to operate with long pulses at lower peak power. This paves the way for future developments such as multiplexing emitters with optical combiners at higher optical pump power levels.

Furthermore, the frequency-to-time distribution in the emitted THz pulses offers exciting opportunities for new applications, particularly in THz ranging, where time-domain techniques like FMCW detection can be utilized. The broad bandwidth of the THz pulses and the linear frequency evolution over time also open doors for enhanced spectroscopy and imaging applications.

**Funding**. This work was supported by the IdEx University of Bordeaux / Grand Research Program LIGHT (contract ANR 10-IDEX-0003) and TERAPPY Project (contract ANR-24-CE42-2479).

**Data availability.** Data underlying the results presented in this paper are not publicly available at this time but may be obtained from the authors upon reasonable request.

**Disclosures**. The authors declare no conflicts of interest.